\documentclass{article}%
\usepackage{amssymb}
\usepackage{amsmath}
\usepackage{hyperref}
\usepackage{amsfonts}
\usepackage{graphicx}%
\providecommand{\U}[1]{\protect\rule{.1in}{.1in}}
\begin{document}

\title{The Conducting Ring Viewed as a Wormhole}
\author{T Curtright$^{1}$\thanks{{\small curtright@miami.edu}}, H Alshal$^{1,2}%
$\thanks{{\small alshal@cu.edu.eg}}, P Baral$^{1}$, S Huang$^{1}$, J Liu$^{1}%
$, K Tamang$^{1}$, X Zhang$^{1}$, and Y Zhang$^{1}$\\$^{1}$Department of Physics, University of Miami, Coral Gables, FL 33124-8046, USA\\$^{2}$Department of Physics, Cairo University, Giza, 12613, Egypt}
\date{11 May 2018}
\maketitle

\begin{abstract}
We compute the exterior Green function for a grounded equi-potential circular
ring in two-dimensional electrostatics by treating the system geometrically as
a \textquotedblleft squashed wormhole\textquotedblright\ with an image charge
located in a novel but obvious position, thereby implementing a method first
suggested in 1897 by Sommerfeld. \ We compare and contrast the strength and
location of the image charge in the wormhole picture with that of the
conventional point of view where an image charge is located inside the
circular ring. \ While the two viewpoints give mathematically equivalent Green
functions, we believe they provide strikingly different physics perspectives.
\ We also comment on earlier Green function results by Hobson in 1900, and by
Davis and Reitz in 1971, who applied Sommerfeld's method to analyze a grounded
conducting circular disk in three-dimensional electrostatics.

\end{abstract}

\begin{center}
\vfill%

{\includegraphics[
trim=0.000000in 0.000000in 0.012069in 0.000000in,
height=2.1395in,
width=1.6181in
]%
{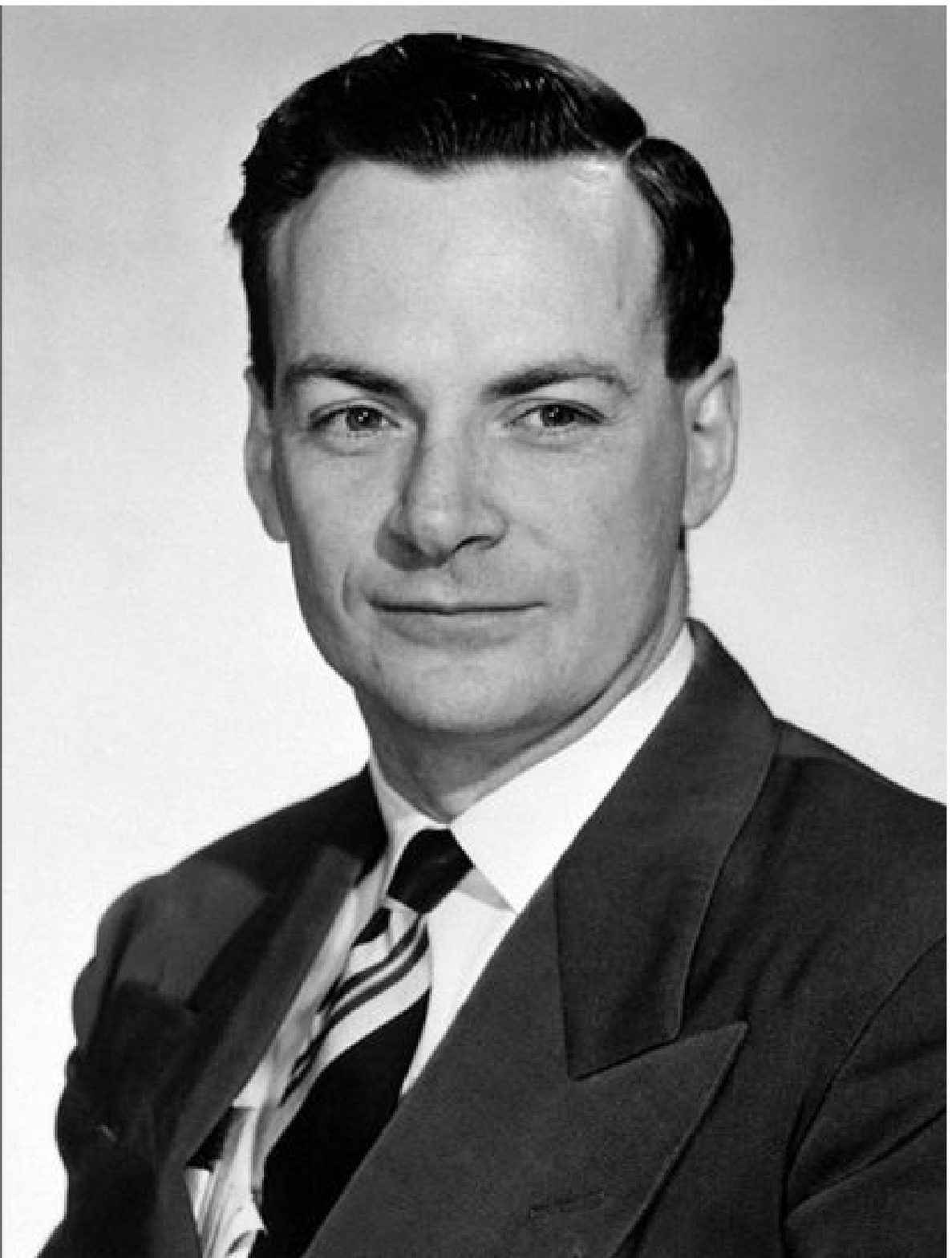}%
}
\ \ \ \ \
{\includegraphics[
trim=0.000000in 0.000000in 0.010524in 0.000000in,
height=2.1777in,
width=1.8597in
]%
{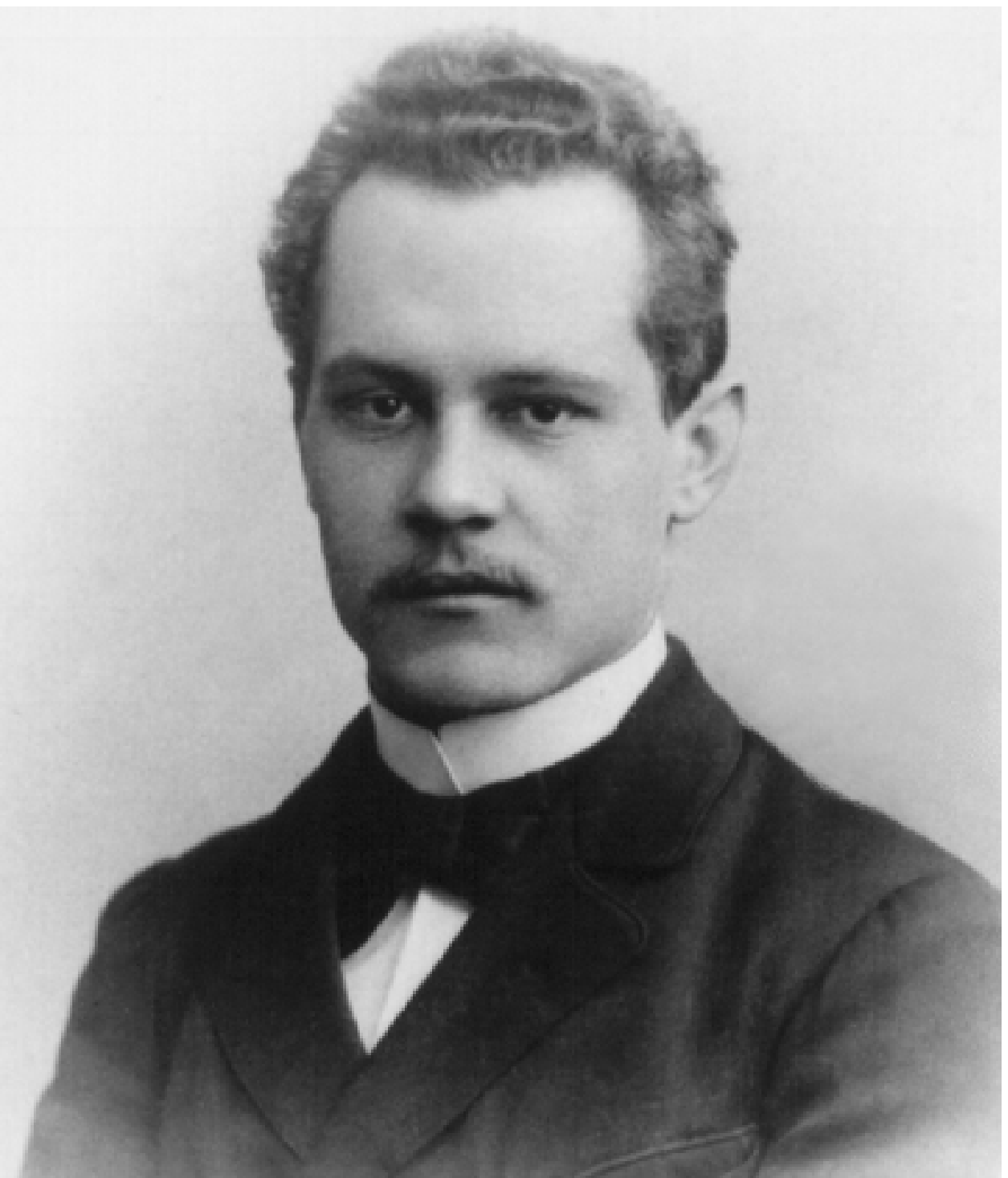}%
}
\bigskip

Richard Feynman (1918-1988) and Arnold Sommerfeld (1868-1951)

\vfill

\end{center}

\section*{Introduction}

\href{https://en.wikipedia.org/wiki/Richard_Feynman}{Richard Feynman}
constantly emphasized that insight could be gained by approaching a problem
from a different point of view \cite{Mehra}. \ We follow that philosophy here
to construct the Green function \textquotedblleft$G_{o}$\textquotedblright%
\ for a grounded circular conducting ring in two-dimensional (2D)
electrostatics by stressing the geometrical aspects of a method first employed
in the late 19th century by
\href{https://en.wikipedia.org/wiki/Arnold_Sommerfeld}{Arnold Sommerfeld}
\cite{Eckert}, albeit not for this specific problem. \ Although the method was
introduced some 120 years ago \cite{Sommerfeld}, we believe it suggests
insights that are not widely appreciated. \ Indeed, while Sommerfeld's method
has occasionally been successfully employed during the intervening century to
solve a handful of otherwise difficult electrostatic problems
\cite{DavisReitz,Hobson}, we believe that a stronger emphasis on its
geometrical aspects is worthwhile and justifies applying the method to a
broader class of problems, even those which are not difficult to solve by
other means. \ To that end we reconsider the grounded circular ring.

Although the Green function with homogeneous Dirichlet boundary conditions is
well-known for this problem and can be obtained by several methods, here we
compute $G_{o}$ via a novel method that uses \textquotedblleft
wormholes\textquotedblright\ \cite{MorrisThorne,JamesEtAl}\ in a 2D Riemannian
space. \ We begin by considering a 2D version of the classic Ellis wormhole
\cite{Ellis}, whose circular slices have radii given by $r\left(  w\right)
=\sqrt{R^{2}+w^{2}}$ where $R$ is a constant and $w$ is the standard real
parameterization of the surface, with $-\infty\leq w\leq+\infty$, constructed
so that the region near $w=0$ is a curved bridge\ \cite{EinsteinRosen}\ that
connects two distinct branches of the surface, with those branches approaching
two separate flat planes asymptotically as $w\rightarrow\pm\infty$ (see the Figures).

We build appropriate Green functions for the invariant Laplacian $\nabla^{2}$
acting on this geometry, initially by imposing boundary conditions only as
$w\rightarrow\pm\infty$, to obtain \textquotedblleft$G$\textquotedblright, and
finally by requiring that the Green function also vanish at $w=0$, i.e. at
radius $R$, to obtain $G_{o}$. \ 

The latter \textquotedblleft grounded\textquotedblright\ Green function
$G_{o}$ is constructed by placing a source and its negative image at exactly
the same radius and in precisely the same direction on the surface, but on
opposite branches. \ 

Next we introduce a deformation of the Ellis wormhole whose radii are given by
the $p$-norm $r\left(  w\right)  =\left(  R^{p}+\left\vert w\right\vert
^{p}\right)  ^{1/p}$. \ We compute Green functions $G$ and $G_{o}$ for this
family of surfaces as well. \ Finally, we consider the $p\rightarrow1$ limit
of this family of surfaces and Green functions. \ 

In this $p\rightarrow1$ limit, where $r\left(  w\right)  =R+\left\vert
w\right\vert $ is the so-called Manhattan\ norm, both branches of the surface
are completely \textquotedblleft squashed flat\textquotedblright. \ That is to
say, the surface degenerates into two distinct flat planes joined together
along the circumference of a circular \textquotedblleft
doorway\textquotedblright, i.e. a ring of radius $R$ (for views of the surface
as $p$ approaches $1$, see the Figures). \ However, the interior of the ring
is \emph{excluded} from either plane. \ Moreover, in this limit it is clear
that by restricting $G_{o}$ to have source and field points on just one of the
distinct planes, a Green function for the grounded conducting ring in 2D is
obtained. \ Minor rearrangements of the terms shows that $G_{o}$ is in exact
agreement with the usual symmetrized Green function for the grounded ring, as
obtained by considering only \emph{one} Euclidean plane with an image charge
placed inside the ring. \bigskip

Generalizations to several problems in higher dimensions are forthcoming
\cite{AlshalCurtright}.\bigskip

\section*{Electrostatics in two Euclidean dimensions}

The point-particle electric potential on the 2D Euclidean plane,
$\mathbb{E}_{2}$, is well-known to be logarithmic. \ For a unit point charge
located at the origin, up to a constant $R$ that sets the distance scale,
\begin{equation}
\Phi_{\mathbb{E}_{2}}\left(  \overrightarrow{r}\right)  =-\frac{1}{2\pi}%
\ln\left(  r/R\right)  \ ,\text{\ \ \ \ \ }\nabla^{2}\Phi_{\mathbb{E}_{2}%
}\left(  \overrightarrow{r}\right)  =-~\delta^{2}\left(  \overrightarrow{r}%
\right)  \ , \label{PointOnPlane}%
\end{equation}
Consequently, a Green function for the plane is
\begin{equation}
G_{\mathbb{E}_{2}}\left(  \overrightarrow{r_{1}};\overrightarrow{r_{2}%
}\right)  =-\frac{1}{2\pi}\ln\left(  \left\vert \overrightarrow{r_{1}%
}-\overrightarrow{r_{2}}\right\vert /R\right)  =-\frac{1}{4\pi}\ln\left(
\frac{r_{1}^{2}+r_{2}^{2}-2r_{1}r_{2}\cos\left(  \theta_{1}-\theta_{2}\right)
}{R^{2}}\right)  \ ,\ \ \ \ \ \nabla^{2}G_{\mathbb{E}_{2}}\left(
\overrightarrow{r_{1}};\overrightarrow{r_{2}}\right)  =-~\delta^{2}\left(
\overrightarrow{r_{1}}-\overrightarrow{r_{2}}\right)  \label{PlaneGreen}%
\end{equation}
This result is translationally invariant on the plane, that is to say,
$G_{\mathbb{E}_{2}}\left(  \overrightarrow{r_{1}};\overrightarrow{r_{2}%
}\right)  $ depends only on the difference $\overrightarrow{r_{1}%
}-\overrightarrow{r_{2}}$. \ Moreover, this choice for the Green function
incorporates boundary conditions at spatial infinity that mimic the behavior
in (\ref{PointOnPlane}). \ So any sufficiently localized\footnote{For example,
it is sufficient but not necessary that all the moments of the charge are
finite, i.e. $\int r^{m}\rho\left(  \overrightarrow{r}\right)  ~d^{2}r<\infty$
for all positive integer $m$.} charge distribution $\rho\left(
\overrightarrow{r}\right)  $ on the plane gives rise to the usual linear
superposition,
\begin{equation}
\Phi\left(  \overrightarrow{r_{1}}\right)  =\int G_{\mathbb{E}_{2}}\left(
\overrightarrow{r_{1}};\overrightarrow{r_{2}}\right)  ~\rho\left(
\overrightarrow{r_{2}}\right)  ~d^{2}r_{2} \label{LinearSuperposition}%
\end{equation}
where we assume the integral is well-defined and finite. \ For a localized
charge distribution, $\Phi\left(  \overrightarrow{r_{1}}\right)
\underset{r_{1}\rightarrow\infty}{\sim}-\frac{Q}{2\pi}~\ln\left(
r_{1}/R\right)  +O\left(  1/r_{1}\right)  $ where $Q=\int\rho\left(
\overrightarrow{r_{2}}\right)  ~d^{2}r_{2}$ is the total charge.

\section*{Electrostatics in two curved dimensions}

On a Riemannian manifold, described by a metric $g_{\mu\nu}$, distance
increments are given by $ds$ where%
\begin{equation}
\left(  ds\right)  ^{2}=g_{\mu\nu}~dx^{\mu}dx^{\nu}%
\end{equation}
Summation over integer $\mu$ and $\nu$ is implicitly understood, with
$1\leq\mu,\nu\leq N$ for an $N$-dimensional manifold. \ The invariant
Laplacian on such a manifold is given by
\begin{equation}
\nabla^{2}=\frac{1}{\sqrt{g}}~\partial_{\mu}\left(  \sqrt{g}~g^{\mu\nu
}\partial_{\nu}\right)
\end{equation}
where $g\equiv\det g_{\mu\nu}$ and $g^{\mu\nu}$ is the matrix inverse of
$g_{\mu\nu}$, written using the standard \textquotedblleft raised
index\textquotedblright\ notation. \ So, with an implicit sum over $\mu$, we
have $g^{\lambda\mu}g_{\mu\nu}=\delta_{\ \nu}^{\lambda}=1$ if $\lambda=\nu$
but $0$ otherwise, i.e. $\delta_{\ \nu}^{\lambda}$ is the unit matrix. \ 

Consider now an infinite 2D manifold, e.g. an unbounded surface of revolution
embedded in 3D, with%
\begin{equation}
\left(  ds\right)  ^{2}=\left(  dw\right)  ^{2}+r^{2}\left(  w\right)  \left(
d\theta\right)  ^{2} \label{General2DWormhole}%
\end{equation}
The variables $w$ and $\theta$ take on values $-\infty\leq w\leq+\infty$ and
$0\leq\theta\leq2\pi$, and $r\left(  w\right)  $ is assumed to be a positive,
non-vanishing \textquotedblleft radius\textquotedblright\ function that has a
minimum at $w=0$ and becomes infinite as $w\rightarrow\pm\infty$. \ For a
fixed angle, radial displacements on the manifold are determined just by
$ds=\pm dw$. \ 

We refer to $w>0$ and $w<0$ respectively as the \textquotedblleft
upper\textquotedblright\ and \textquotedblleft lower\textquotedblright%
\ branches of the manifold, and we call the region near $w=0$ the
\textquotedblleft bridge\textquotedblright\ between the two branches. \ For
visualization purposes, see the examples shown in the Figures.

The metric, its inverse, and its determinant for this manifold are given by%
\begin{equation}
g_{ww}=1\ ,\ \ \ g_{\theta\theta}=r^{2}\left(  w\right)  \ ,\ \ \ g^{ww}%
=1\ ,\ \ \ g^{\theta\theta}=\frac{1}{r^{2}\left(  w\right)  }\ ,\ \ \ g=r^{2}%
\left(  w\right)
\end{equation}
For the particular cases of interest to follow, $r\left(  w\right)  $ is
symmetric under $w\rightarrow-w$ and monotonic in $\left\vert w\right\vert $.
\ In any case, the form of the metric leads to%
\begin{equation}
\nabla^{2}=\frac{1}{g}\left(  r\left(  w\right)  \frac{\partial}{\partial
w}\left(  r\left(  w\right)  \frac{\partial}{\partial w}\right)
+\frac{\partial^{2}}{\partial\theta^{2}}\right)
\end{equation}
which is more transparent after changing radial variable to
\begin{equation}
u=\int_{0}^{w}\frac{d\varpi}{r\left(  \varpi\right)  }%
\end{equation}

In terms of $u$ and $\theta$ it is now easy to find harmonic functions, i.e.
solutions of $\nabla^{2}h=0$, since%
\begin{equation}
\nabla^{2}=\frac{1}{g}\left(  \frac{\partial^{2}}{\partial u^{2}}%
+\frac{\partial^{2}}{\partial\theta^{2}}\right)
\end{equation}
For fixed radius, those $h$ which are periodic in $\theta$ with period $2\pi$
therefore have the form%
\begin{equation}
h_{0}\left(  u\right)  =a+b~u\text{ \ \ and \ \ }\ h_{m}^{\pm}\left(
u,\theta\right)  =c_{m}~e^{-mu}e^{\pm im\theta}\text{ \ \ for any integer
}m\gtrless0
\end{equation}
Note that $h_{0}$\ is isotropic, i.e. it has no $\theta$ dependence.
\ Reverting back to the original variables for the 2D manifold
(\ref{General2DWormhole}), these periodic harmonic functions are given by \
\begin{equation}
h_{0}\left(  w,\theta\right)  =a+b\int_{0}^{w}\frac{d\varpi}{r\left(
\varpi\right)  }\ ,\ \ \ h_{m}^{\pm}\left(  w,\theta\right)  =c_{m}~e^{\pm
im\theta}~\exp\left(  -m\int_{0}^{w}\frac{d\varpi}{r\left(  \varpi\right)
}\right)  \text{ \ \ for any integer }m\gtrless0 \label{2DManifoldHarmonics}%
\end{equation}

By using these harmonic functions, it is not difficult to build Green
functions for the invariant Laplacian on the manifold, as solutions to
\begin{equation}
\nabla^{2}G\left(  w_{1},\theta_{1};w_{2},\theta_{2}\right)  =-\frac{1}%
{\sqrt{g}}~\delta\left(  w_{1}-w_{2}\right)  ~\delta\left(  \theta_{1}%
-\theta_{2}\right)  \label{GreenFunctionEquation}%
\end{equation}
A Green function symmetric under $\left(  w_{1},\theta_{1}\right)
\leftrightarrow\left(  w_{2},\theta_{2}\right)  $ is given by%
\begin{align}
G\left(  w_{1},\theta_{1};w_{2},\theta_{2}\right)   &  =-\frac{1}{4\pi}\left(
\int_{w_{<}}^{w_{>}}\frac{d\varpi}{r\left(  \varpi\right)  }\right)  +\frac
{1}{2\pi}\sum_{m=1}^{\infty}\frac{1}{m}~e^{-m\int_{w_{<}}^{w_{>}}\frac
{d\varpi}{r\left(  \varpi\right)  }}\cos\left(  m\left(  \theta_{1}-\theta
_{2}\right)  \right) \nonumber\\
&  =-\frac{1}{4\pi}\left(  \int_{w_{<}}^{w_{>}}\frac{d\varpi}{r\left(
\varpi\right)  }\right)  -\frac{1}{4\pi}\ln\left(  1+e^{-2\int_{w_{<}}^{w_{>}%
}\frac{d\varpi}{r\left(  \varpi\right)  }}-2e^{-\int_{w_{<}}^{w_{>}}%
\frac{d\varpi}{r\left(  \varpi\right)  }}\cos\left(  \theta_{1}-\theta
_{2}\right)  \right)  \label{GreenWormhole}%
\end{align}
where $w_{>,<}=\max,\min\left(  w_{1},w_{2}\right)  $. \ In the second line of
(\ref{GreenWormhole}), we have summed the Taylor series using%
\begin{equation}
\sum_{n=1}^{\infty}\frac{1}{n}~z^{n}\cos\left(  n\theta\right)  =-\frac{1}%
{2}\ln\left(  1+z^{2}-2z\cos\theta\right)
\end{equation}
$G\left(  w_{1},\theta_{1};w_{2},\theta_{2}\right)  $ has the usual
interpretation as the potential at $\left(  w_{1},\theta_{1}\right)  $ due to
a point source at $\left(  w_{2},\theta_{2}\right)  $.

If $\int_{w_{<}}^{w_{>}}\frac{du}{r\left(  u\right)  }\rightarrow+\infty$
either as $w_{>}\rightarrow+\infty$ or as $w_{<}\rightarrow-\infty$, the Green
function (\ref{GreenWormhole}) satisfies the boundary condition that all the
$\theta$-dependent terms vanish as either $w_{1}\rightarrow\infty$ for fixed
$w_{2}$ or as $w_{2}\rightarrow-\infty$ for fixed $w_{1}$, leaving only the
isotropic term which diverges in either of those limits. \ However, when
$w_{2}=0$ the value of $G$ is not so obvious. \ Nevertheless, a simple linear
combination can be taken to construct a Green function that vanishes at
$w_{2}=0$, namely,%
\begin{equation}
G_{o}\left(  w_{1},\theta_{1};w_{2},\theta_{2}\right)  =G\left(  w_{1}%
,\theta_{1};w_{2},\theta_{2}\right)  -G\left(  w_{1},\theta_{1};-w_{2}%
,\theta_{2}\right)  \label{GreenGroundedWormhole}%
\end{equation}
Note that $G_{o}$ is manifestly an odd function of $w_{2}$ and is also an odd
function of $w_{1}$ as a consequence of $G\left(  w_{1},\theta_{1}%
;w_{2},\theta_{2}\right)  =G\left(  w_{2},\theta_{2};w_{1},\theta_{1}\right)
$.

This $G_{o}$ has a simple interpretation as the potential at $\left(
w_{1},\theta_{1}\right)  $ due to a point charge source at $\left(
w_{2},\theta_{2}\right)  $ and a negative image charge of that point source at
$\left(  -w_{2},\theta_{2}\right)  $. \ The source and image charges are
therefore of \emph{equal} magnitude but opposite sign, and are positioned in
an \emph{obvious} way on \emph{opposite} branches of the manifold. \ So, if
both $w_{1}$ and $w_{2}$ are restricted to one branch of the manifold,
(\ref{GreenFunctionEquation}) will still hold on that branch, since the image
will produce an additional Dirac delta only on the other branch. \ Therefore
$G_{o}$ is an appropriate Green function to solve $\nabla^{2}\Phi=-\rho$ on
the upper branch of the manifold with the condition $\Phi=0$ on the inner
boundary of that branch, i.e. on the circular ring at $w=0$.

To verify that $G$ is indeed a solution to (\ref{GreenFunctionEquation}) two
steps are required. \ Firstly, it is necessary to show $\nabla^{2}G=0$ for
non-coincident points on the manifold, but this is guaranteed by the
harmonicity of the functions used to build $G$. \ Secondly, it is also
necessary to show how the Dirac deltas arise. \ 

In regard to this second step, note the discontinuities in the radial
derivatives%
\begin{equation}
r\left(  w_{1}\right)  \frac{d}{dw_{1}}\int_{w_{<}}^{w_{>}}\frac{d\varpi
}{r\left(  \varpi\right)  }=\Theta\left(  w_{1}-w_{2}\right)  -\Theta\left(
w_{2}-w_{1}\right)
\end{equation}%
\begin{equation}
r\left(  w_{1}\right)  \frac{d}{dw_{1}}\exp\left(  -m\int_{w_{<}}^{w_{>}}%
\frac{d\varpi}{r\left(  \varpi\right)  }\right)  =-m\left(  \Theta\left(
w_{1}-w_{2}\right)  -\Theta\left(  w_{2}-w_{1}\right)  \right)  \exp\left(
-m\int_{w_{<}}^{w_{>}}\frac{d\varpi}{r\left(  \varpi\right)  }\right)
\end{equation}
where the Heaviside step function is $\Theta\left(  z\right)  =1$ if $z>0$ or
$\Theta\left(  z\right)  =0$ if $z<0$, and whose derivative is a Dirac delta,
$d\Theta\left(  z\right)  /dz=\delta\left(  z\right)  $. \ Thus%
\begin{align}
\frac{d}{dw_{1}}\left(  r\left(  w_{1}\right)  \frac{d}{dw_{1}}\int_{w_{<}%
}^{w_{>}}\frac{d\varpi}{r\left(  \varpi\right)  }\right)   &  =2~\delta\left(
w_{1}-w_{2}\right) \\
\frac{d}{dw_{1}}\left(  r\left(  w_{1}\right)  \frac{d}{dw_{1}}\exp\left(
-m\int_{w_{<}}^{w_{>}}\frac{d\varpi}{r\left(  \varpi\right)  }\right)
\right)   &  =-2m~\delta\left(  w_{1}-w_{2}\right)  +m^{2}\exp\left(
-m\int_{w_{<}}^{w_{>}}\frac{d\varpi}{r\left(  \varpi\right)  }\right)
\end{align}
Using these facts, along with%
\begin{equation}
\frac{1}{2\pi}\sum_{m=-\infty}^{\infty}e^{im\left(  \theta_{1}-\theta
_{2}\right)  }=\delta\left(  \theta_{1}-\theta_{2}\right)
\end{equation}
when acting on a space of periodic functions, the Dirac deltas on the RHS of
(\ref{GreenFunctionEquation}) are verified.

\section*{$p$-norm wormholes}

Next, consider a class of \textquotedblleft$p$-norm radial
functions\textquotedblright\ with $R$ a constant radius, $p\geq1$ a real
number, and
\begin{equation}
r\left(  w\right)  =\left(  R^{p}+\left\vert w\right\vert ^{p}\right)  ^{1/p}
\label{pNormRadius}%
\end{equation}
Particular cases of the 2D manifolds defined by (\ref{pNormRadius}) and
(\ref{General2DWormhole}) are shown in the Figures,\ as 3D embeddings of
surfaces of revolution about the $z$-axis. \ For the special case of an Ellis
wormhole \cite{Ellis}
\begin{equation}
r\left(  w\right)  =\sqrt{R^{2}+w^{2}}%
\end{equation}
In this special case (\ref{GreenWormhole}) becomes%
\begin{gather}
G\left(  w_{1},\theta_{1};w_{2},\theta_{2}\right)  =\frac{1}{4\pi}\ln\left[
\left(  \sqrt{\frac{w_{>}^{2}}{R^{2}}+1}-\frac{w_{>}}{R}\right)  \left(
\sqrt{\frac{w_{<}^{2}}{R^{2}}+1}+\frac{w_{<}}{R}\right)  \right]
\label{EllisG}\\
-\frac{1}{4\pi}\ln\left[  1+\left(  \sqrt{\frac{w_{>}^{2}}{R^{2}}+1}%
-\frac{w_{>}}{R}\right)  ^{2}\left(  \sqrt{\frac{w_{<}^{2}}{R^{2}}+1}%
+\frac{w_{<}}{R}\right)  ^{2}-2\left(  \sqrt{\frac{w_{>}^{2}}{R^{2}}+1}%
-\frac{w_{>}}{R}\right)  \left(  \sqrt{\frac{w_{<}^{2}}{R^{2}}+1}+\frac{w_{<}%
}{R}\right)  \cos\left(  \theta_{1}-\theta_{2}\right)  \right] \nonumber
\end{gather}
while $G_{o}$ is the previous linear combination (\ref{GreenGroundedWormhole}).

A contour plot of $G$ versus $w_{1}$ and $\theta_{1}$ is shown in Figure 10,
for the Ellis wormhole, with unit source at $\left(  w_{2},\theta_{2}\right)
=\left(  1,0\right)  $. \ For comparison, a similar plot of $G_{o}$ is shown
in Figure 11 with the same source as well as its negative image at $\left(
-w_{2},\theta_{2}\right)  =\left(  -1,0\right)  $.

\section*{Squashing the wormhole}

Flattening the two branches of the wormhole, to obtain a \textquotedblleft
two-sided\textquotedblright\ surface of zero \textquotedblleft
height\textquotedblright\ containing a hole of fixed radius $R$, is achieved
by taking the $p=1$ \textquotedblleft Manhattan norm\textquotedblright%
\footnote{In the opposite extreme, where $p\rightarrow\infty$, the $p$-norm
wormhole becomes the right-circular cylinder of \cite{JamesEtAl}, p 488, Eqn
(3).}\
\begin{equation}
r\left(  w\right)  =R+\left\vert w\right\vert
\end{equation}
It suffices to consider two cases for the source and field point locations,
either with $w_{1}$ and $w_{2}$ on the same branch of the manifold (i.e. on
the same side of the two-sided surface), or with $w_{1}$ and $w_{2}$ on
opposite branches (i.e. on opposite sides of the two-sided surface). \ Suppose
$w_{1}$ is always on the upper branch, say. \ Then the two cases lead to $\ $%
\begin{align}
\int_{w_{<}}^{w_{>}}\frac{dv}{R+\left\vert \nu\right\vert }  &  =\ln\left(
\frac{R+w_{>}}{R+w_{<}}\right)  \text{ \ \ if }w_{1}>0\text{ and }w_{2}>0\\
\int_{w_{<}}^{w_{>}}\frac{dv}{R+\left\vert \nu\right\vert }  &  =\int%
_{0}^{w_{1}}\frac{dv}{R+v}+\int_{w_{2}}^{0}\frac{dv}{R-v}=\ln\left(
\frac{R+w_{1}}{R}\right)  +\ln\left(  \frac{R-w_{2}}{R}\right)  \text{ \ \ if
}w_{1}>0\text{ but }w_{2}<0
\end{align}
In the first case, with\ $w_{1}>0$ and $w_{2}>0$, (\ref{GreenWormhole})
becomes%
\begin{align}
\left.  G\left(  w_{1},\theta_{1};w_{2},\theta_{2}\right)  \right\vert
_{\substack{w_{1}>0\\w_{2}>0}}  &  =-\frac{1}{4\pi}\ln\left[  \left(
\frac{R+w_{>}}{R+w_{<}}\right)  \left(  1+\left(  \frac{R+w_{<}}{R+w_{>}%
}\right)  ^{2}-2\left(  \frac{R+w_{<}}{R+w_{>}}\right)  \cos\left(  \theta
_{1}-\theta_{2}\right)  \right)  \right] \nonumber\\
&  =-\frac{1}{4\pi}\ln\left(  \frac{r_{1}^{2}+r_{2}^{2}-2r_{1}r_{2}\cos\left(
\theta_{1}-\theta_{2}\right)  }{r_{1}r_{2}}\right)  \label{Gflat++}%
\end{align}
where in the second line we have identified $r_{1}=R+\left\vert w_{1}%
\right\vert $ and $r_{2}=R+\left\vert w_{2}\right\vert $, to obtain a
symmetrical form that can also be used if $w_{1}<0$ and $w_{2}<0$. \ 

In the second case, with $w_{1}>0$ but $w_{2}<0$, (\ref{GreenWormhole})
becomes
\begin{align}
\left.  G\left(  w_{1},\theta_{1};w_{2},\theta_{2}\right)  \right\vert
_{\substack{w_{1}>0\\w_{2}<0}}  &  =-\frac{1}{4\pi}\ln\left[  \left(
\frac{R+w_{1}}{R}\frac{R-w_{2}}{R}\right)  \left(  1+\left(  \frac{R^{2}%
}{\left(  R+w_{1}\right)  \left(  R-w_{2}\right)  }\right)  ^{2}-\frac{2R^{2}%
}{\left(  R+w_{1}\right)  \left(  R-w_{2}\right)  }\cos\left(  \theta
_{1}-\theta_{2}\right)  \right)  \right] \nonumber\\
&  =-\frac{1}{4\pi}\ln\left(  \frac{r_{1}r_{2}}{R^{2}}+\frac{R^{2}}{r_{1}%
r_{2}}-2\cos\left(  \theta_{1}-\theta_{2}\right)  \right)  \label{Gflat+-}%
\end{align}
where in the last line we have again identified $r_{1}=R+\left\vert
w_{1}\right\vert $ and $r_{2}=R+\left\vert w_{2}\right\vert $, to obtain a
symmetrical form that can also be used if $w_{1}<0$ and $w_{2}>0$. \ Taken
together, (\ref{Gflat++}) and (\ref{Gflat+-}) yield\ $G\left(  w_{1}%
,\theta_{1};w_{2},\theta_{2}\right)  $ for any $w_{1}$ and $w_{2}$. \ 

A contour plot of $G$ versus $w_{1}$ and $\theta_{1}$ for the squashed
wormhole is shown in Figure 12, with unit source at $\left(  w_{2},\theta
_{2}\right)  =\left(  1,0\right)  $. \ 

With these results in hand, it is instructive to compare the result
(\ref{Gflat++}) for the squashed wormhole, with its two branches, to the Green
function for a single copy of the entire Euclidean plane, (\ref{PlaneGreen}),
at identical points. \ The difference is given by%
\begin{equation}
\left.  G\left(  w_{1},\theta_{1};w_{2},\theta_{2}\right)  \right\vert
_{\substack{w_{1}>0\\w_{2}>0}}-G_{\mathbb{E}_{2}}\left(  \overrightarrow{r_{1}%
};\overrightarrow{r_{2}}\right)  =-\frac{1}{4\pi}\ln\left(  \frac{R^{2}}%
{r_{1}r_{2}}\right)  \label{1stDifference}%
\end{equation}
It is also instructive to compare (\ref{Gflat+-}) to (\ref{PlaneGreen}) when
the source point\ for the latter is interior to the circle of radius $R$, at a
location $\overrightarrow{\mathfrak{r}_{2}}=\frac{R^{2}}{r_{2}^{2}%
}~\overrightarrow{r_{2}}$ with $r_{2}>R$, so that $\mathfrak{r}_{2}%
=\frac{R^{2}}{r_{2}}<R$. \ The difference is now given by%
\begin{equation}
\left.  G\left(  w_{1},\theta_{1};w_{2},\theta_{2}\right)  \right\vert
_{\substack{w_{1}>0\\w_{2}<0}}-G_{\mathbb{E}_{2}}\left(  \overrightarrow{r_{1}%
};\frac{R^{2}}{r_{2}^{2}}~\overrightarrow{r_{2}}\right)  =\frac{1}{4\pi}%
\ln\left(  \frac{r_{1}}{r_{2}}\right)  \label{2ndDifference}%
\end{equation}
Note that the terms on the RHS's of (\ref{1stDifference}) and
(\ref{2ndDifference}) are harmonic on $\mathbb{E}_{2}$ so long as $r_{1}%
\neq0\neq r_{2}$, i.e. on the \textquotedblleft punctured
plane\textquotedblright\ these terms are harmonic.

Upon taking the linear combination (\ref{GreenGroundedWormhole}), a Green
function is obtained for the situation where the circular ring connecting the
two flat branches of the squashed manifold is grounded. \ Explicitly, using
(\ref{Gflat++}) and (\ref{Gflat+-}), when $w_{1}$ and $w_{2}$ have the same
sign the result is
\begin{equation}
G_{o}\left(  w_{1},\theta_{1};w_{2},\theta_{2}\right)  =-\frac{1}{4\pi}%
\ln\left(  \frac{1}{r_{1}r_{2}}\left(  r_{1}^{2}+r_{2}^{2}-2r_{1}r_{2}%
\cos\left(  \theta_{1}-\theta_{2}\right)  \right)  \right)  +\frac{1}{4\pi}%
\ln\left(  \frac{r_{1}r_{2}}{R^{2}}+\frac{R^{2}}{r_{1}r_{2}}-2\cos\left(
\theta_{1}-\theta_{2}\right)  \right)
\end{equation}
with $r_{1}=R+\left\vert w_{1}\right\vert $ and $r_{2}=R+\left\vert
w_{2}\right\vert $. \ When $w_{1}$ and $w_{2}$ have opposite signs, $G_{o}$
has an overall sign change because $G_{o}$ is an odd function of $w_{1}$ for
fixed $w_{2}$. \ Combining the various terms then gives the final result for
the squashed wormhole grounded on the circle of radius $R$. \ When $w_{1}$ and
$w_{2}$ have the same sign,%
\begin{equation}
G_{o}\left(  w_{1},\theta_{1};w_{2},\theta_{2}\right)  =-\frac{1}{4\pi}%
\ln\left(  R^{2}\frac{r_{1}^{2}+r_{2}^{2}-2r_{1}r_{2}\cos\left(  \theta
_{1}-\theta_{2}\right)  }{r_{1}^{2}r_{2}^{2}+R^{4}-2R^{2}r_{1}r_{2}\cos\left(
\theta_{1}-\theta_{2}\right)  }\right)  \label{Goflat}%
\end{equation}
As a function of $\overrightarrow{r_{1}}$ and $\overrightarrow{r_{2}}$, this
is precisely the well-known Green function for points outside a grounded ring
on a single copy of the Euclidean plane $\mathbb{E}_{2}$, as obtained by other
means. \ Note that it is symmetric under $\left(  w_{1},\theta_{1}\right)
\leftrightarrow\left(  w_{2},\theta_{2}\right)  $, or equivalently under
$\overrightarrow{r_{1}}\leftrightarrow\overrightarrow{r_{2}}$.

A contour plot of $G_{o}$ for the squashed wormhole is shown in Figure 13,
again with a unit source\ at $\left(  w_{2},\theta_{2}\right)  =\left(
1,0\right)  $ as well as its negative image at $\left(  -w_{2},\theta
_{2}\right)  =\left(  -1,0\right)  $.

Curiously, however, there is a subtlety due to a slight disparity between
(\ref{Goflat}) and the Euclidean Green function obtained just by taking the
linear combination (LC)
\begin{equation}
G_{\mathbb{E}_{2}\text{LC}}\left(  \overrightarrow{r_{1}}%
;\overrightarrow{r_{2}}\right)  =G_{\mathbb{E}_{2}}\left(
\overrightarrow{r_{1}};\overrightarrow{r_{2}}\right)  -G_{\mathbb{E}_{2}%
}\left(  \overrightarrow{r_{1}};\frac{R^{2}}{r_{2}^{2}}~\overrightarrow{r_{2}%
}\right)
\end{equation}
This combination can be interpreted as the potential at the field point
$\overrightarrow{r_{1}}$ due to a unit source charge outside the ring, at
position $\overrightarrow{r_{2}}$, and a negative unit image source inside the
ring, at position $\frac{R^{2}}{r_{2}^{2}}~\overrightarrow{r_{2}}$. \ The
disparity is given by%
\begin{equation}
G_{o}\left(  w_{1},\theta_{1};w_{2},\theta_{2}\right)  =G_{\mathbb{E}%
_{2}\text{LC}}\left(  \overrightarrow{r_{1}};\overrightarrow{r_{2}}\right)
-\frac{1}{2\pi}\ln\left(  \frac{R}{r_{2}}\right)  \label{Disparity}%
\end{equation}
and in part it arises because, unlike $G_{o}$, the combined $G_{\mathbb{E}%
_{2}\text{LC}}$ is \emph{not} symmetric under $\overrightarrow{r_{1}%
}\leftrightarrow\overrightarrow{r_{2}}$. \ But perhaps a better way to
understand the disparity is to construct the Green function for a grounded
hypersphere in $N$ spatial dimensions and then take the limit as
$N\rightarrow2$ (for example, see \cite{AlshalCurtright}). \ In any case, the
additional log term on the RHS of (\ref{Disparity}) is harmonic on the
punctured plane. \ Consequently, both of the differential equations
$\nabla_{r_{1}}^{2}G\left(  \overrightarrow{r_{1}};\overrightarrow{r_{2}%
}\right)  =-~\delta^{2}\left(  \overrightarrow{r_{1}}-\overrightarrow{r_{2}%
}\right)  $ and $\nabla_{r_{2}}^{2}G\left(  \overrightarrow{r_{1}%
};\overrightarrow{r_{2}}\right)  =-~\delta^{2}\left(  \overrightarrow{r_{1}%
}-\overrightarrow{r_{2}}\right)  $ will still be satisfied for points
$\overrightarrow{r_{1}}$ and $\overrightarrow{r_{2}}$ outside the circle of
radius $R$, as will the homogeneous Dirichlet boundary condition for all
$\overrightarrow{r_{2}}$ on the circle, namely, $\left.  G\left(
\overrightarrow{r_{1}};\overrightarrow{r_{2}}\right)  \right\vert
_{\overrightarrow{r_{2}}=R\widehat{r}}=0$, no matter whether $G$ is taken to
be $G_{o}$ or $G_{\mathbb{E}_{2}\text{LC}}$.

\section*{Relating image charge distributions by inversion}

Coordinate inversion on the plane maps the interior of the circular ring to
the exterior, and vice versa, and therefore inversion should be expected to
relate the standard image method, where the so-called
\href{https://en.wikipedia.org/wiki/William_Thomson,_1st_Baron_Kelvin}{Kelvin}
image is placed inside the ring, to the Sommerfeld method for the squashed
wormhole. \ Indeed, inversion of the source position is the technique that is
normally invoked to locate Kelvin images for grounded sphere Green functions
in any dimension. \ 

The inversion mapping is defined by%
\begin{equation}
\overrightarrow{\mathfrak{r}}=\frac{R^{2}}{r^{2}}~\overrightarrow{r}%
\end{equation}
Radial distances change under the inversion, $\mathfrak{r}=R^{2}/r$, but
angles do not, $\widehat{\mathfrak{r}}=\widehat{r}$. \ Under an inversion the
Laplacian does \emph{not} transform into a geometric factor multiplying just
the Laplacian, \emph{except in 2D}. \ In other dimensions, $N$, the Laplacian
mixes with the \emph{scale operator} under an inversion, as follows (for
example, see \cite{AlshalCurtright}).%
\begin{equation}
\nabla_{r}^{2}=\left(  \frac{\mathfrak{r}^{2}}{R^{2}}\right)  ^{2}\left(
\nabla_{\mathfrak{r}}^{2}+\frac{2\left(  2-N\right)  }{\mathfrak{r}^{2}%
}~D_{\mathfrak{r}}\right)  \ ,\ \ \ \ \ D_{\mathfrak{r}}%
=\overrightarrow{\mathfrak{r}}\cdot\overrightarrow{\nabla}_{\mathfrak{r}}
\label{InversionOnLaplacian}%
\end{equation}
This statement may be understood by considering harmonic functions, upon
noting that\ under inversions $r$ effectively becomes $1/r$, and only in 2D do
both $r^{l}$ and $r^{-l}$ appear as factors in harmonic functions. \ In $N$
dimensions the factors are $r^{l}$ and $r^{2-N-l}$.

Moreover, $G$ itself is \emph{not} invariant under the inversion, \emph{except
in 2D}. \ This is obvious on dimensional grounds, since $G\left(
\overrightarrow{r},0\right)  \propto1/r^{N-2}\underset{\text{inversion}%
}{\longrightarrow}\mathfrak{r}^{N-2}/R^{2N-4}\propto\left(  \mathfrak{r}%
^{2N-4}/R^{2N-4}\right)  G\left(  \overrightarrow{\mathfrak{r}},0\right)  $.
\ More precisely, in $N$ spatial dimensions,%
\begin{equation}
G\left(  \overrightarrow{r_{1}};\overrightarrow{r_{2}}\right)
\underset{\text{inversion}}{\longrightarrow}\frac{\mathfrak{r}_{1}%
^{N-2}~\mathfrak{r}_{2}^{N-2}}{R^{2N-4}}~G\left(  \overrightarrow{\mathfrak{r}%
_{1}};\overrightarrow{\mathfrak{r}_{2}}\right)
\end{equation}
Note the symmetry under $\overrightarrow{r_{1}}\leftrightarrow
\overrightarrow{r_{2}}$\ is maintained under $\overrightarrow{\mathfrak{r}%
_{1}}\leftrightarrow\overrightarrow{\mathfrak{r}_{2}}$. \ The complete
transformation of the differential equation for the Green function in $N$
dimensions is%
\begin{align}
\nabla_{r_{1}}^{2}G\left(  \overrightarrow{r_{1}};\overrightarrow{r_{2}%
}\right)   &  =-\frac{1}{\sqrt{g}}~\delta^{N}\left(  \overrightarrow{r_{1}%
}-\overrightarrow{r_{2}}\right)  \underset{\text{inversion}}{\longrightarrow
}\\
\left(  \frac{\mathfrak{r}_{1}^{2}}{R^{2}}\right)  ^{2}\left(  \nabla
_{\mathfrak{r}_{1}}^{2}+\frac{2\left(  2-N\right)  }{\mathfrak{r}_{1}^{2}%
}~D_{\mathfrak{r}_{1}}\right)  \left(  \frac{\mathfrak{r}_{1}^{N-2}%
~\mathfrak{r}_{2}^{N-2}}{R^{2N-4}}~G\left(  \overrightarrow{\mathfrak{r}_{1}%
};\overrightarrow{\mathfrak{r}_{2}}\right)  \right)   &  =-\frac
{\mathfrak{r}_{1}^{N+1}}{R^{2N}}~\delta\left(  \mathfrak{r}_{1}-\mathfrak{r}%
_{2}\right)  ~\delta^{N-1}\left(  \widehat{\mathfrak{r}}_{1}%
-\widehat{\mathfrak{r}}_{2}\right)
\end{align}
Again the 2D case is especially simple. \ In 2D, up to a common factor, the
equation is unchanged in form by the inversion. \ Thus the following must
simultaneously hold in 2D.
\begin{equation}
\nabla_{r_{1}}^{2}G\left(  \overrightarrow{r_{1}};\overrightarrow{r_{2}%
}\right)  =-\frac{1}{r_{1}}~\delta\left(  r_{1}-r_{2}\right)  ~\delta\left(
\theta_{1}-\theta_{2}\right)  \ ,\ \ \ \nabla_{\mathfrak{r}_{1}}^{2}G\left(
\overrightarrow{\mathfrak{r}_{1}};\overrightarrow{\mathfrak{r}_{2}}\right)
=-\frac{1}{\mathfrak{r}_{1}}~\delta\left(  \mathfrak{r}_{1}-\mathfrak{r}%
_{2}\right)  ~\delta\left(  \mathfrak{\theta}_{1}-\mathfrak{\theta}%
_{2}\right)
\end{equation}

In view of these results, it is not difficult to map only the lower branch of
the squashed wormhole into the interior of the ring while leaving the upper
branch unchanged, thereby obtaining a single copy of 2D Euclidean space that
includes both the exterior and the interior of the circle. \ In the course of
this inversion, the image charge is moved to its more conventional position
within the ring. \ We leave the details as an exercise for the reader.

\section*{Brief remarks on the grounded conducting disk in 3D}

\href{https://en.wikipedia.org/wiki/E._W._Hobson}{Hobson} used Sommerfeld's
method and a clever coordinate choice to find the Green function for an
equi-potential circular disk in three Euclidean dimensions (3D) \cite{Hobson}.
\ In this approach, the disk serves as a doorway between two copies of 3D
Euclidean space, $\mathbb{E}_{3}$, with the unit source and field point
located in one copy of $\mathbb{E}_{3}$,\ and an equal strength, negative
image of the source obviously placed in the same position as the source except
in the second copy of $\mathbb{E}_{3}$.

Some seventy years later, Davis and Reitz independently solved the same
problem, again using Sommerfeld's method, with an emphasis on the use of
complex analysis to construct the Green function \cite{DavisReitz}. \ In this
regard, their approach is more in line with Sommerfeld's original analysis,
wherein complex variables also play a central role.

Neither of these treatments invoke Riemannian geometry as we have done here
for the conducting ring in 2D. \ However it is possible in principle to
consider the conducting disk in 3D as a squashed oblate spheroid, and thereby
obtain the Green function for the disk by taking a limit of Green functions on
branched manifolds connected by spheroidal generalizations of the Ellis
wormhole, analogous to the $p$-norm wormholes used above. \ This more
geometrical treatment will be discussed elsewhere \cite{AlshalCurtright}.

\section*{Conclusions}

We have obtained the Green function for a grounded circular ring in two
spatial dimensions by first constructing Green functions on two-dimensional
Riemannian manifolds (commonly but rather unfortunately known as
\textquotedblleft wormholes\textquotedblright) and then by squashing these
manifolds to produce two copies of flat Euclidean planes creased together
along a circular hole of radius $R$. \ The distribution of source and image
charges on the final squashed manifold illustrates Sommerfeld's generalization
of
\href{https://en.wikipedia.org/wiki/William_Thomson,_1st_Baron_Kelvin}{Thomson}%
's method.

Sommerfeld knew that his generalized method could be used to solve a large
variety of problems \cite{Sommerfeld}, writing to
\href{https://en.wikipedia.org/wiki/Felix_Klein}{Klein} in the spring of 1897
(see \cite{Eckert} page 80):

\begin{quote}
\textquotedblleft The number of boundary value problems solvable by means of
my elaborated Thomson's method of images is very great.\textquotedblright
\end{quote}

\noindent But he does not seem to have explicitly pursued this during the next
half-century, perhaps because more interesting mathematics and physics
questions captured his attention.

In our opinion, the most prescient aspect of Sommerfeld's nineteenth century
work lies in its suggestion that physical problems in electromagnetic theory
may be simplified and perhaps more easily understood through the study of
Riemannian geometries, a view that developed much later in general relativity.
\ However, like \href{https://en.wikipedia.org/wiki/Bernhard_Riemann}{Riemann}
before him, in 1897 Sommerfeld had no reason to include time along with the
spatial dimensions of his envisioned manifolds, thus making his work
premature. \ 

Nevertheless, considering its application of Riemann's ideas from geometry and
complex analysis to higher dimensional branched manifolds, we believe
Sommerfeld's work should be recognized as a legitimate precursor to the
wormhole studies that appeared a few decades later
\cite{Flamm,EinsteinRosen,Ellis} and continue to the present day
\cite{MorrisThorne,JamesEtAl}. \ We hope our paper encourages readers to share
this opinion.\bigskip

\noindent\textbf{Acknowledgements} \ It has been our pleasure to reconsider
this elementary subject during the year of the Feynman Centennial and the
Sommerfeld Sesquicentennial. \ This work was supported in part by a University
of Miami Cooper Fellowship, and by a Clark Way Harrison Visiting Professorship
at Washington University in Saint Louis.

\section*{Figures}

The first nine Figures show 3D embeddings of various 2D $p$-norm wormholes,
where
\begin{equation}
\left(  ds\right)  ^{2}=\left(  dw\right)  ^{2}+r^{2}\left(  w\right)  \left(
d\theta\right)  ^{2}=\left(  dx\right)  ^{2}+\left(  dy\right)  ^{2}+\left(
dz\right)  ^{2} \tag{F1}%
\end{equation}%
\begin{equation}
x\left(  w,\theta\right)  =r\left(  w\right)  \cos\theta\ ,\ \ \ y\left(
w,\theta\right)  =r\left(  w\right)  \sin\theta\ ,\ \ \ r\left(  w\right)
=\left(  R^{p}+\left(  w^{2}\right)  ^{p/2}\right)  ^{1/p} \tag{F2}%
\end{equation}%
\begin{equation}
z\left(  w\right)  =\int_{0}^{w}\sqrt{1-\left(  dr\left(  \varpi\right)
/d\varpi\right)  ^{2}}d\varpi=\int_{0}^{w}\sqrt{1-\left(  \varpi^{2}\right)
^{p-1}\left(  R^{p}+\left(  \varpi^{2}\right)  ^{p/2}\right)  ^{\frac{2}{p}%
-2}}\,d\varpi\tag{F3}%
\end{equation}
For example, for $p=2$,%
\begin{equation}
z\left(  w\right)  =R\ln\left(  \frac{w+\sqrt{R^{2}+w^{2}}}{R}\right)
=R\operatorname{arcsinh}\left(  \frac{w}{R}\right)  \tag{F4}%
\end{equation}
For generic $p$, it is easiest to obtain $z\left(  w\right)  $ by numerical
solution of
\begin{equation}
\frac{dz\left(  w\right)  }{dw}=\sqrt{1-\left(  w^{2}\right)  ^{p-1}\left(
R^{p}+\left(  w^{2}\right)  ^{p/2}\right)  ^{\frac{2}{p}-2}} \tag{F5}%
\end{equation}
with initial condition $z\left(  0\right)  =0$. \ \newpage

\noindent Figures 1-9: \ Embedded surfaces for $p$-norm wormholes, where
$p=1+1/2^{k}$, $k=0,\ 1,\ 2,\ 3,\ 4,\ 5,\ 6,\ 7,$\ $8$. \ All plots are in
units where $R=1$, with $0\leq\theta\leq2\pi$ and $-2\leq w\leq2$. \ Upper and
lower branches of the surfaces are shown in orange and green,
respectively.\bigskip

\noindent\hspace{-0.5in}%
{\parbox[b]{2.5438in}{\begin{center}
\includegraphics[
trim=0.000000in 0.000000in 0.000000in -0.316260in,
height=2.611in,
width=2.5438in
]%
{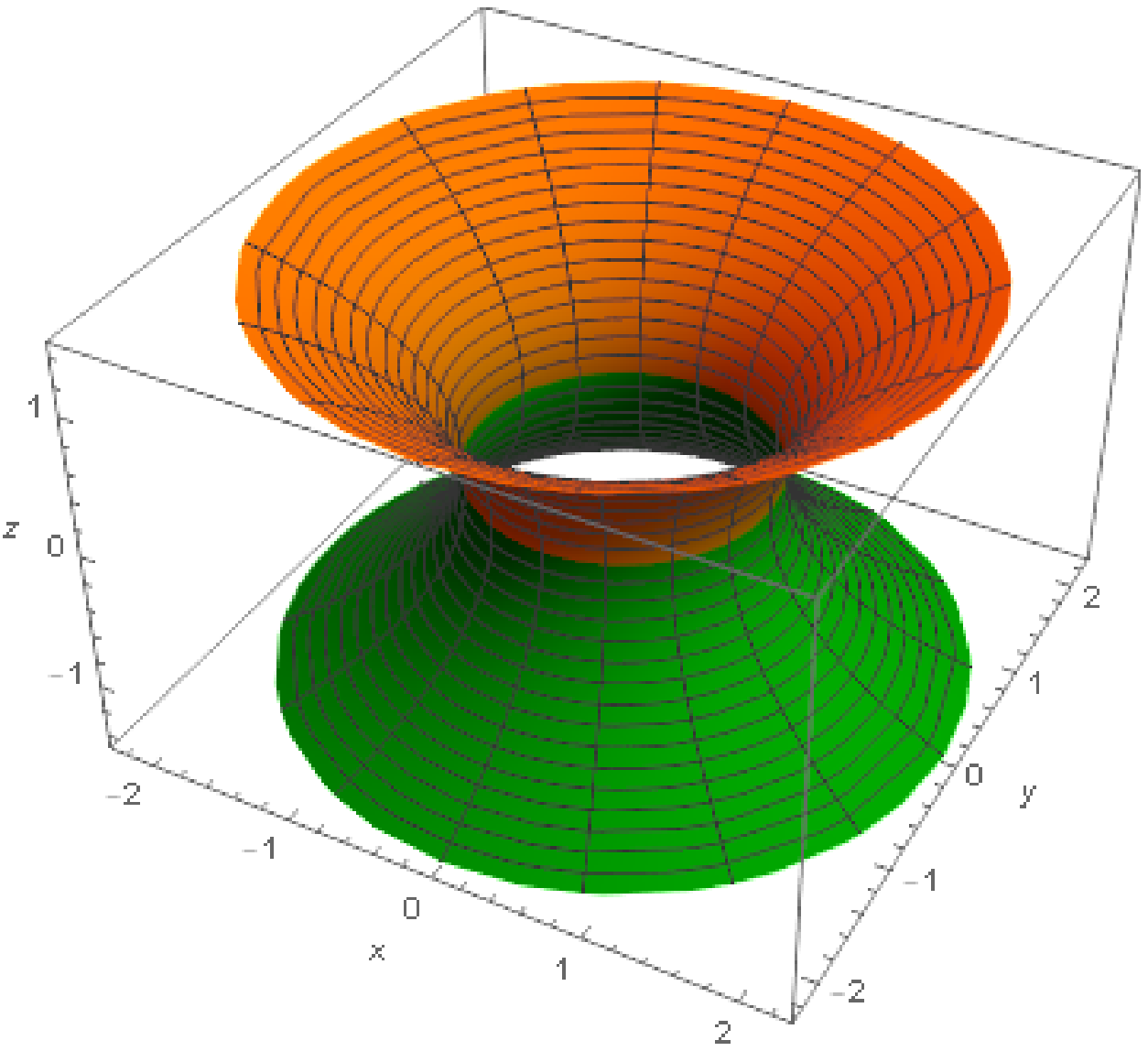}%
\\
Figure 1: $\ p=2$%
\end{center}}}
{\parbox[b]{2.5438in}{\begin{center}
\includegraphics[
trim=0.000000in 0.000000in 0.000000in -0.302285in,
height=2.4284in,
width=2.5438in
]%
{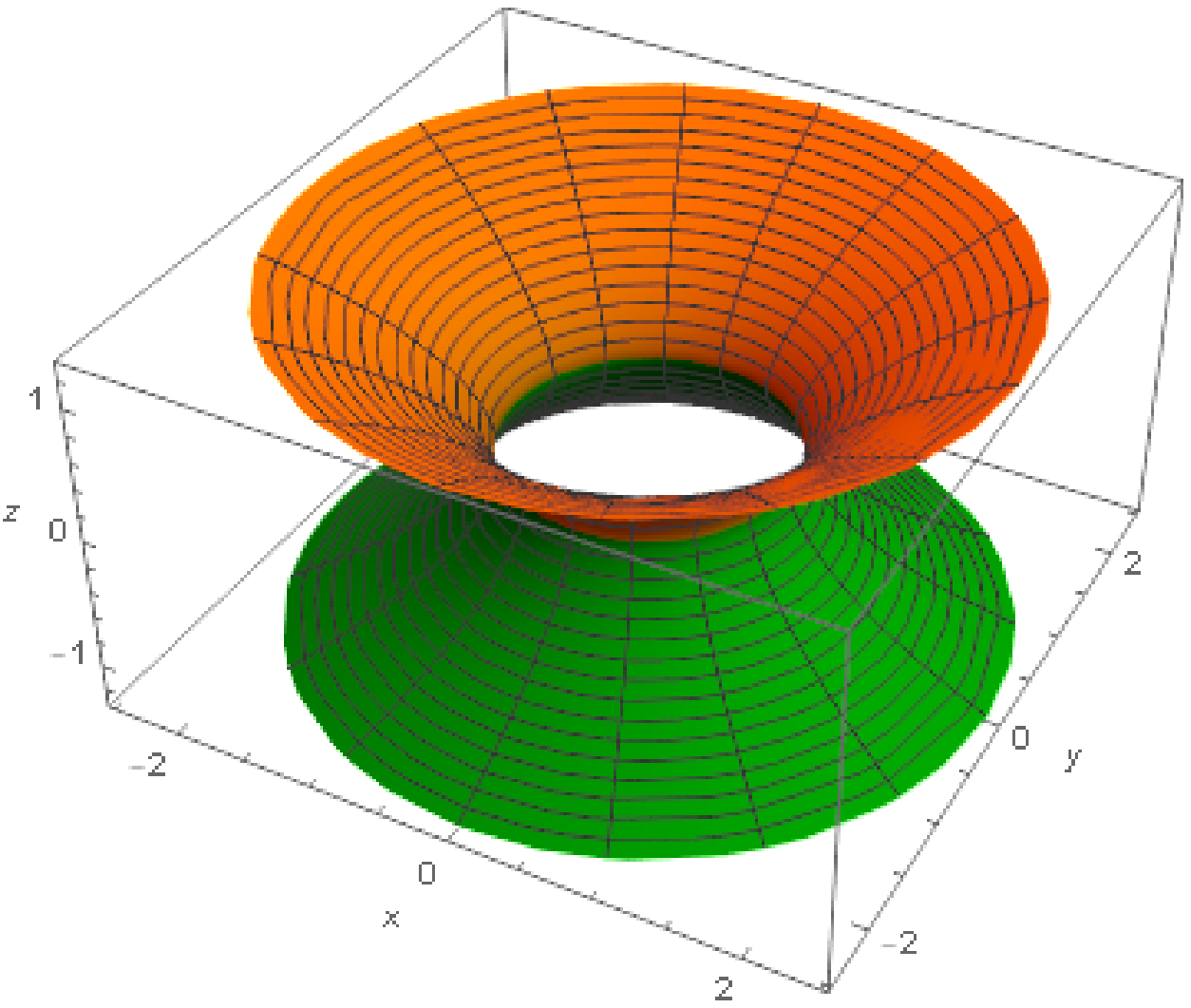}%
\\
Figure 2: $\ p=3/2$%
\end{center}}}
{\parbox[b]{2.5438in}{\begin{center}
\includegraphics[
trim=0.000000in 0.000000in 0.000000in -0.182539in,
height=2.1685in,
width=2.5438in
]%
{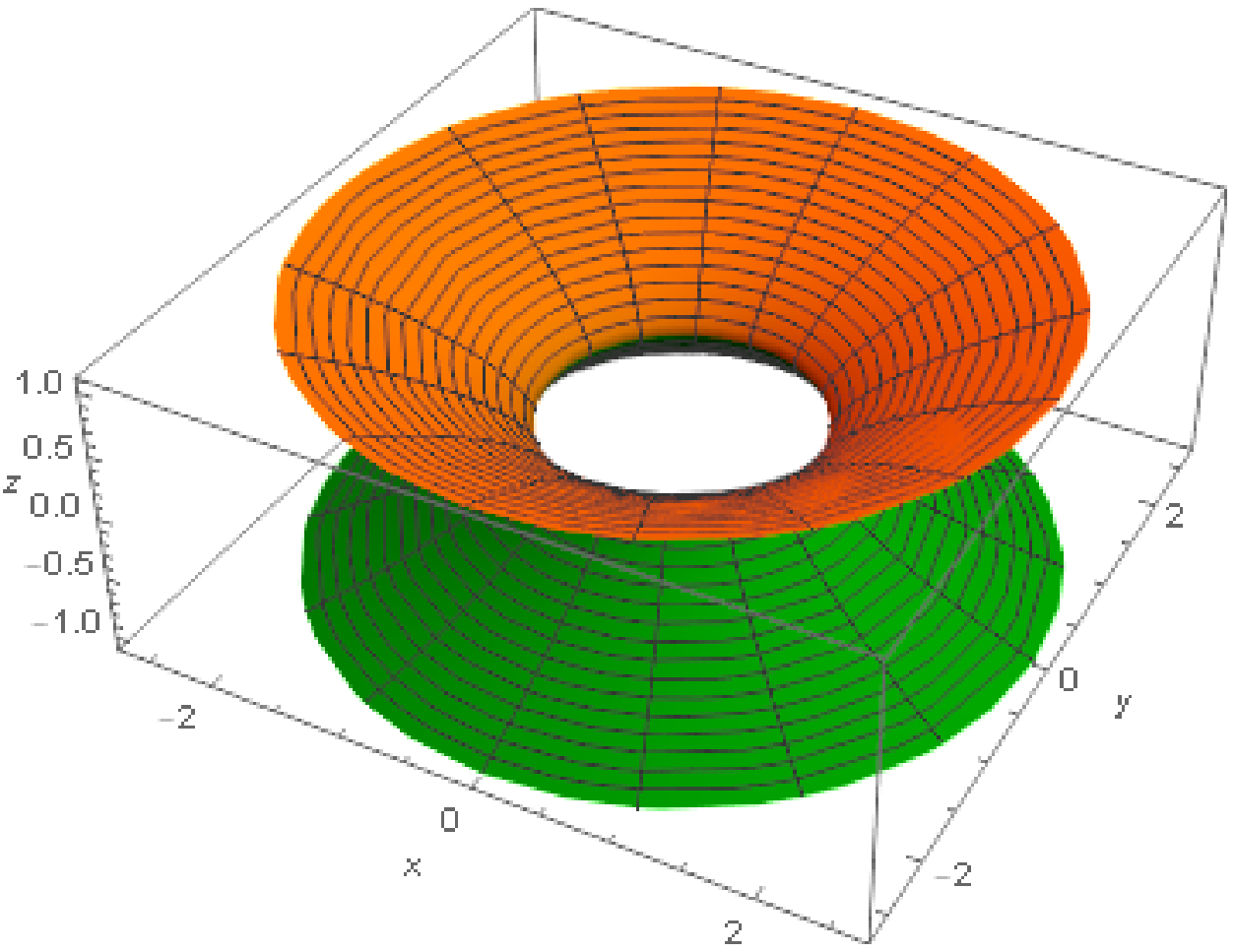}%
\\
Figure 3: $\ p=5/4$%
\end{center}}}

\noindent\hspace{-0.5in}%
{\parbox[b]{2.5438in}{\begin{center}
\includegraphics[
trim=0.000000in 0.000000in 0.000000in -0.185013in,
height=2.0232in,
width=2.5438in
]%
{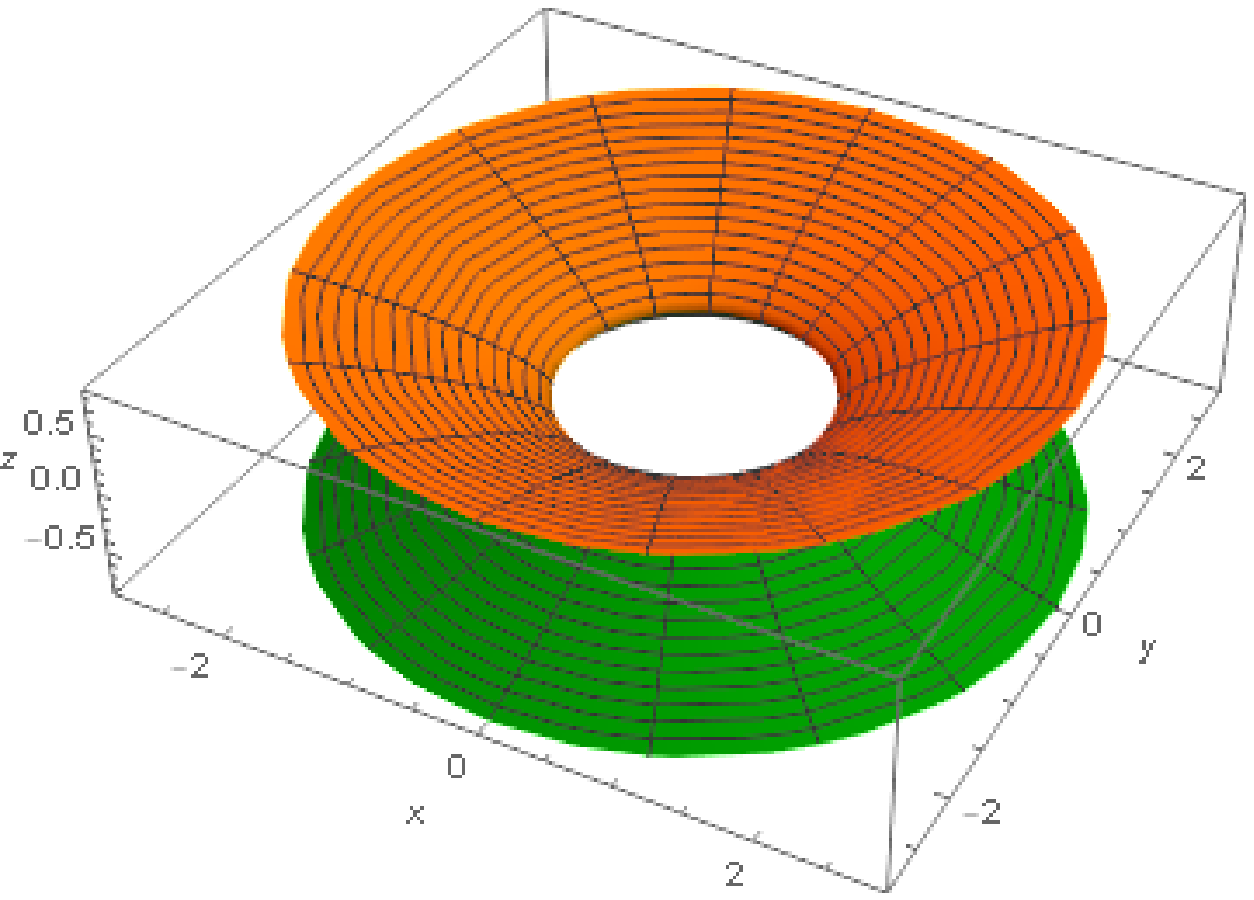}%
\\
Figure 4: $\ p=9/8$%
\end{center}}}
{\parbox[b]{2.5438in}{\begin{center}
\includegraphics[
trim=0.000000in 0.000000in 0.000000in -0.197571in,
height=1.9178in,
width=2.5438in
]%
{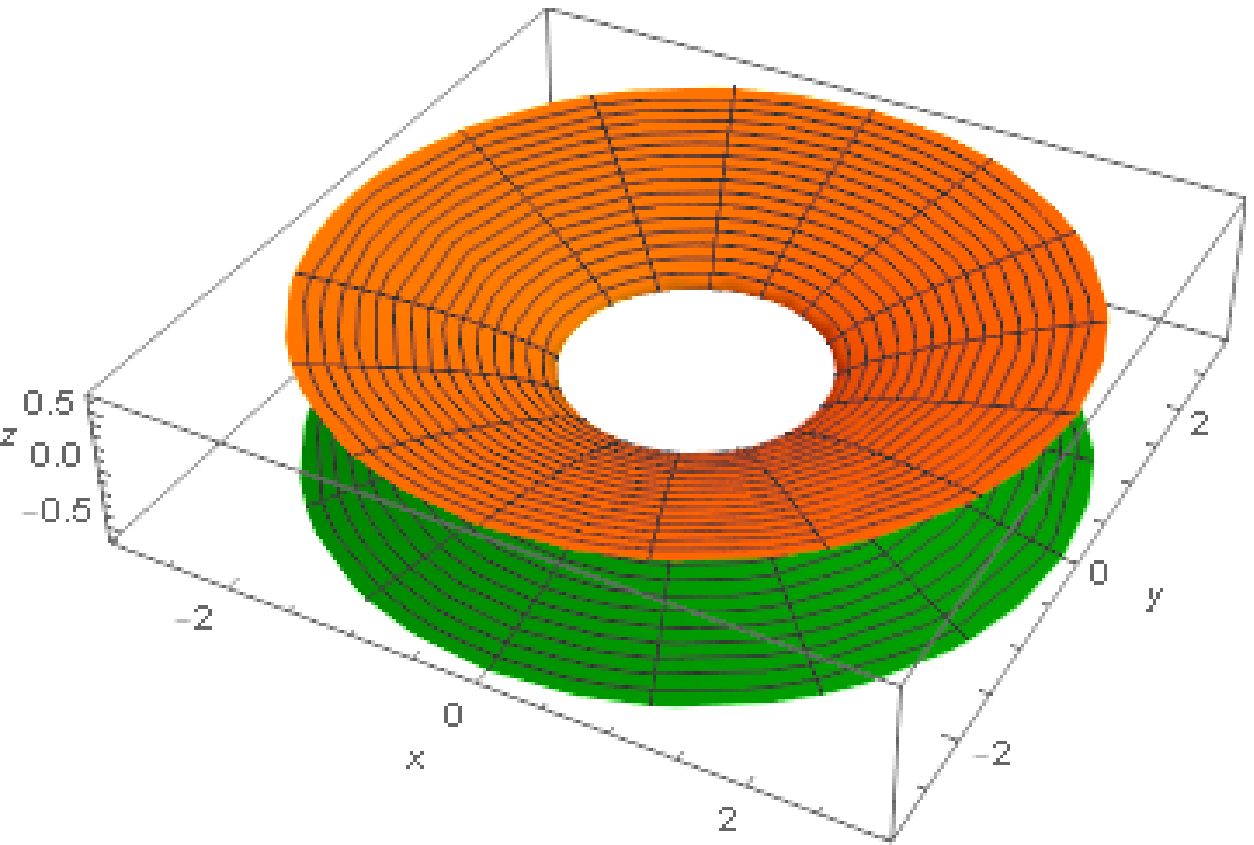}%
\\
Figure 5: $\ p=17/16$%
\end{center}}}
{\parbox[b]{2.5438in}{\begin{center}
\includegraphics[
trim=0.000000in 0.000000in 0.000000in -0.141497in,
height=1.8115in,
width=2.5438in
]%
{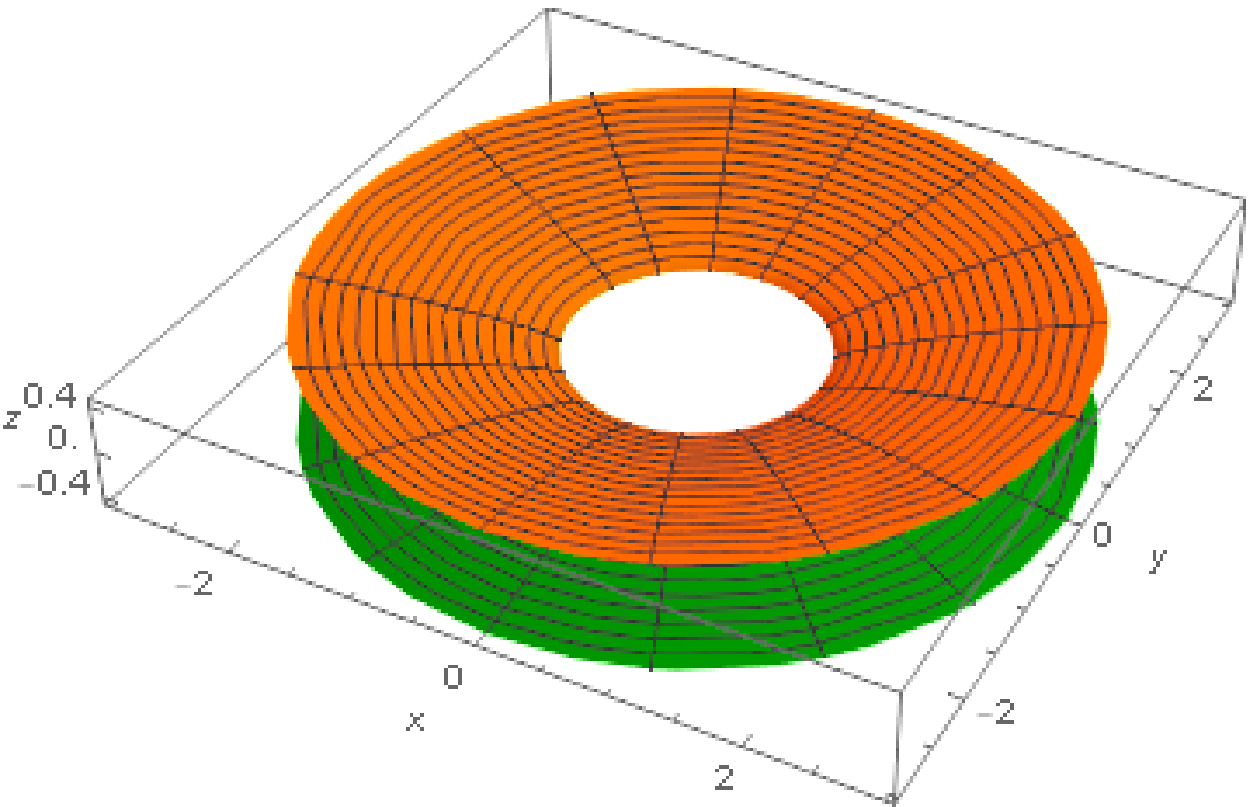}%
\\
Figure 6: $\ p=33/32$%
\end{center}}}

\noindent\hspace{-0.5in}%
{\parbox[b]{2.5438in}{\begin{center}
\includegraphics[
trim=0.000000in 0.000000in 0.000000in -0.121018in,
height=1.7443in,
width=2.5438in
]%
{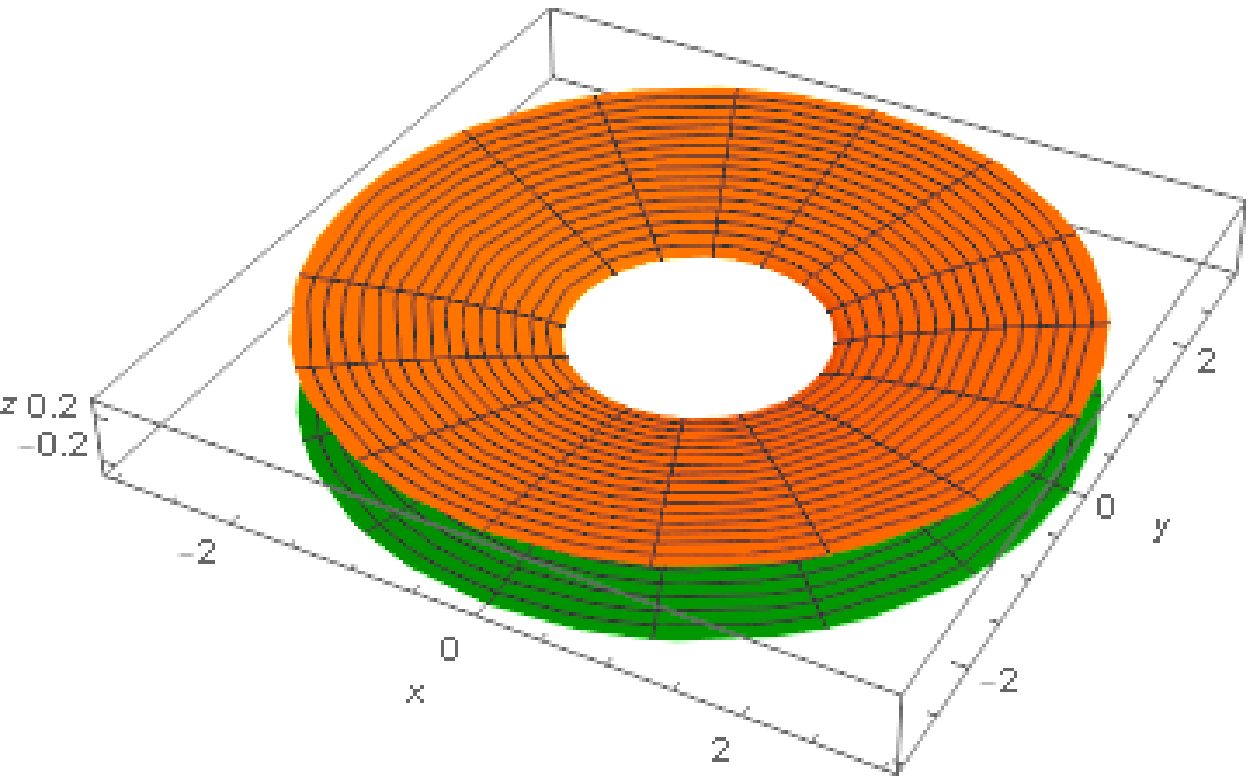}%
\\
Figure 7: $\ p=65/64$%
\end{center}}}
{\parbox[b]{2.5438in}{\begin{center}
\includegraphics[
trim=0.000000in 0.000000in 0.000000in -0.126284in,
height=1.7061in,
width=2.5438in
]%
{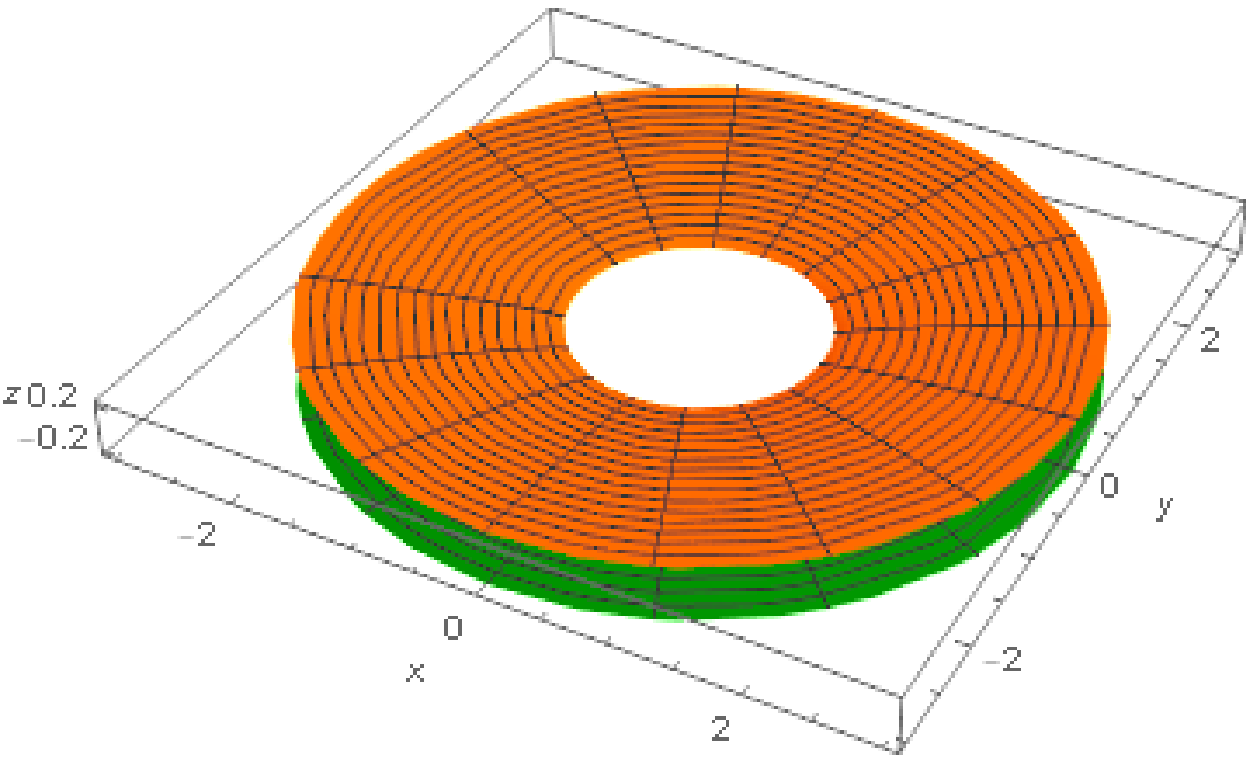}%
\\
Figure 8: $\ p=129/128$%
\end{center}}}
{\parbox[b]{2.5438in}{\begin{center}
\includegraphics[
trim=0.000000in 0.000000in 0.000000in -0.151740in,
height=1.677in,
width=2.5438in
]%
{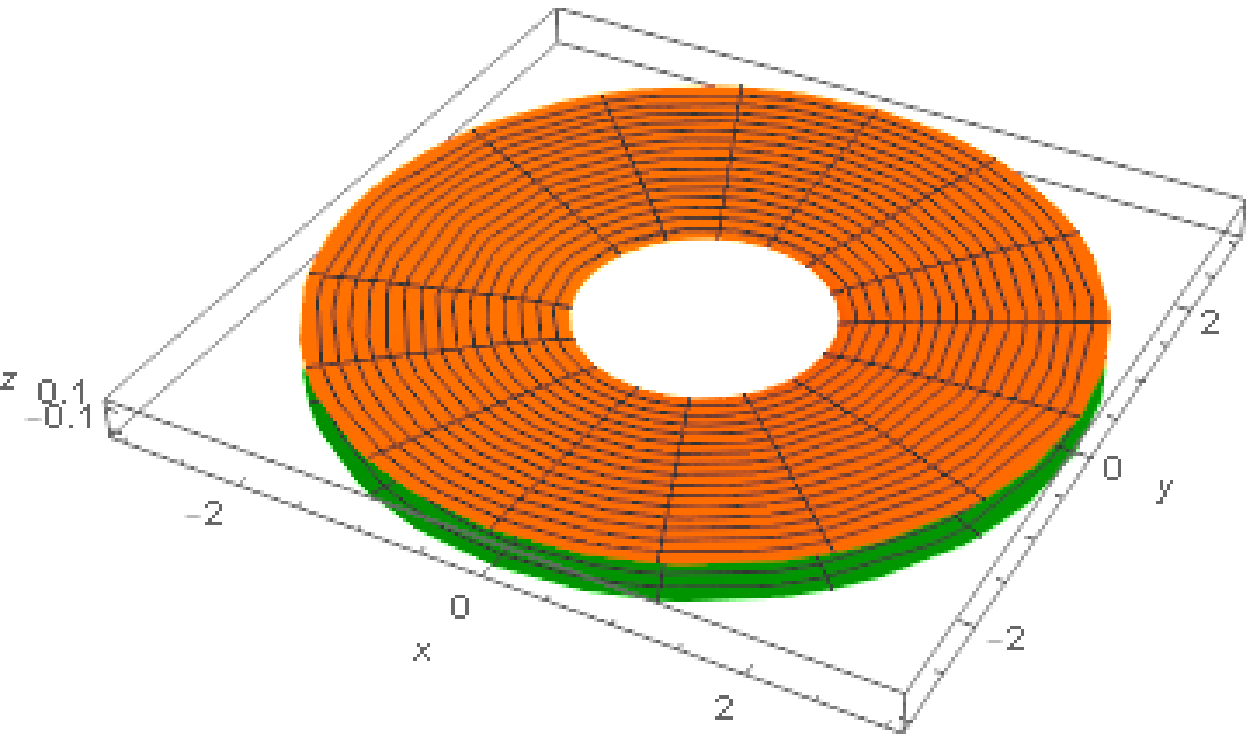}%
\\
Figure 9: $\ p=257/256$%
\end{center}}}

\newpage

\noindent The next four Figures show contour plots of various Green functions,
with $-\pi\leq\theta_{1}\leq\pi$ along the vertical axes, and $-5\leq
w_{1}\leq5$ along the horizontal axes.\bigskip

\noindent Figure 10: \ A plot of (\ref{EllisG}) with source at $\left(
w_{2},\theta_{2}\right)  =\left(  1,0\right)  $.\bigskip

\noindent Figure 11: \ A plot of (\ref{GreenGroundedWormhole}) using
(\ref{EllisG})\ with source at $\left(  w_{2},\theta_{2}\right)  =\left(
1,0\right)  $ \& image at $\left(  -w_{2},\theta_{2}\right)  =\left(
-1,0\right)  $.\bigskip

\noindent Figure 12: \ A plot of (\ref{Gflat++}) and (\ref{Gflat+-}) with
source at $\left(  w_{2},\theta_{2}\right)  =\left(  1,0\right)  $.\bigskip

\noindent Figure 13: \ A plot of (\ref{Goflat}), correctly signed, with source
at $\left(  w_{2},\theta_{2}\right)  =\left(  1,0\right)  $ \& image at
$\left(  -w_{2},\theta_{2}\right)  =\left(  -1,0\right)  $.\bigskip

\begin{center}%
{\parbox[b]{2.5438in}{\begin{center}
\includegraphics[
trim=0.000000in 0.000000in 0.000000in -0.460499in,
height=2.7945in,
width=2.5438in
]%
{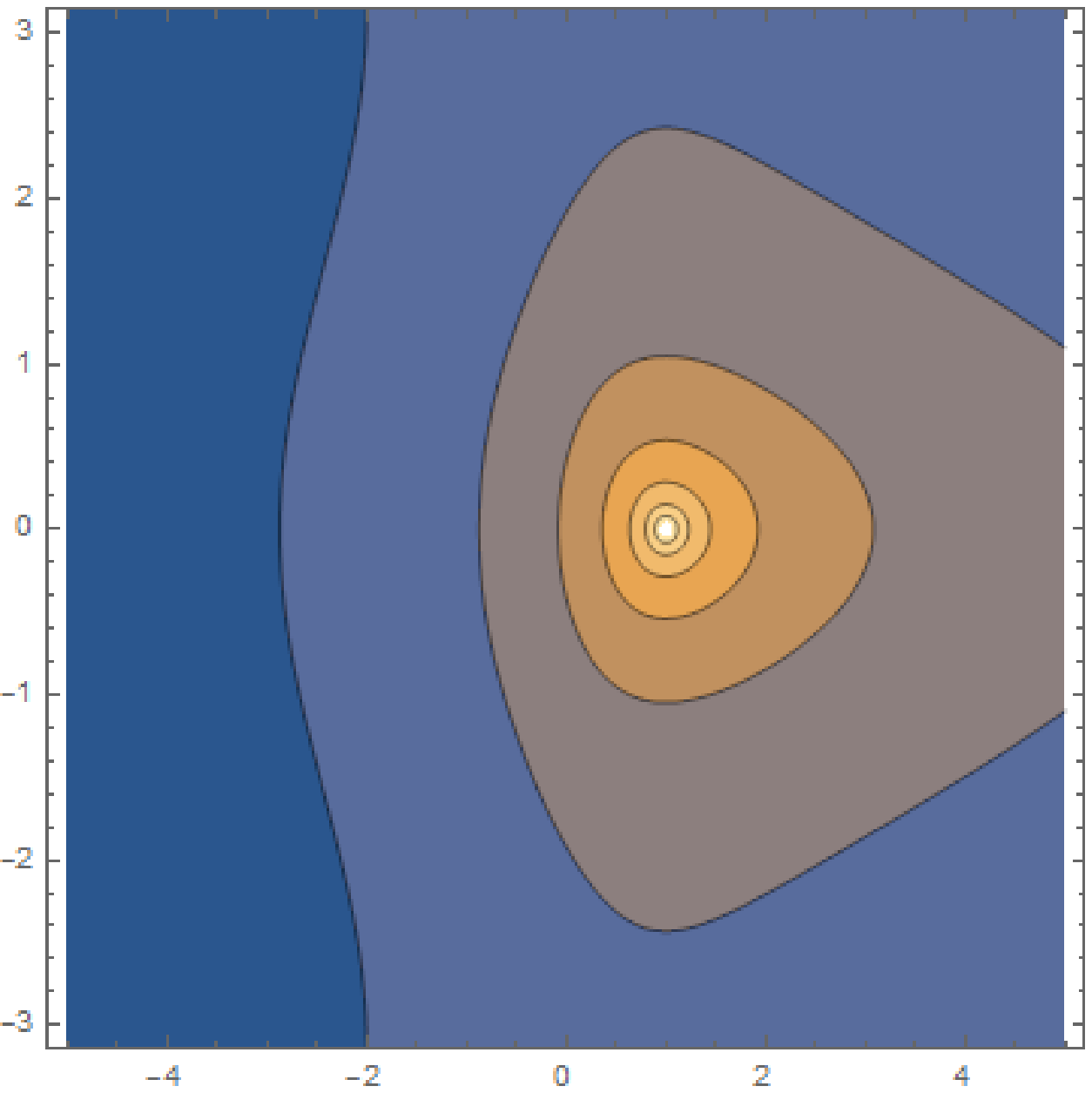}%
\\
Figure 10: \ Contour plot of $G$ for the Ellis wormhole.
\end{center}}}
\ \ \ \
{\parbox[b]{2.5438in}{\begin{center}
\includegraphics[
trim=0.000000in 0.000000in 0.000000in -0.461462in,
height=2.7945in,
width=2.5438in
]%
{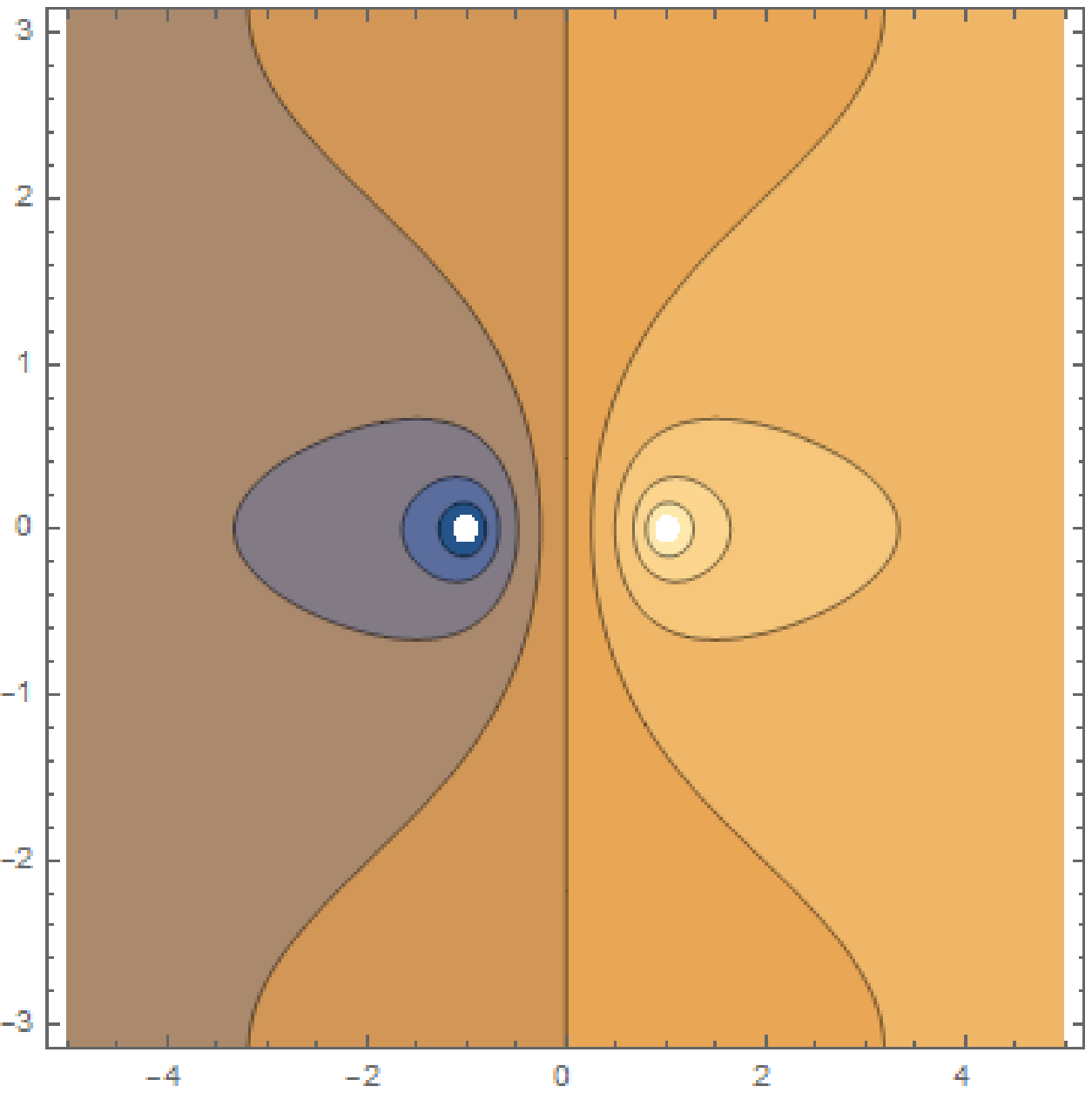}%
\\
Figure 11: $\ $Contour plot of $G_{o}$ for the Ellis wormhole.
\end{center}}}

\bigskip%

{\parbox[b]{2.5438in}{\begin{center}
\includegraphics[
trim=0.000000in 0.000000in 0.000000in -0.461462in,
height=2.7945in,
width=2.5438in
]%
{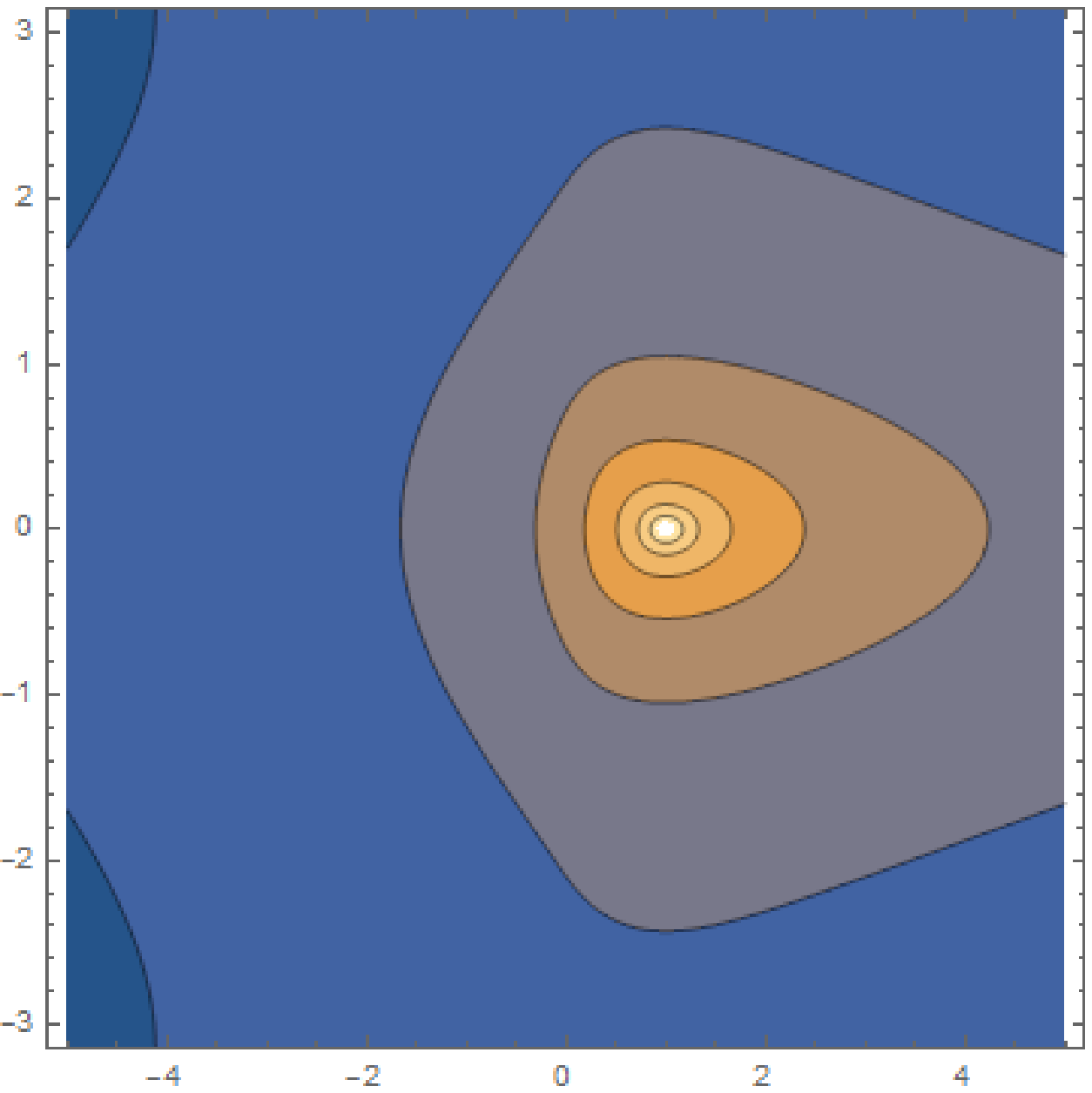}%
\\
Figure 12: $\ $Contour plot of $G$ for the squashed wormhole.
\end{center}}}
\ \ \ \ \
{\parbox[b]{2.5438in}{\begin{center}
\includegraphics[
trim=0.000000in 0.000000in 0.000000in -0.461462in,
height=2.7945in,
width=2.5438in
]%
{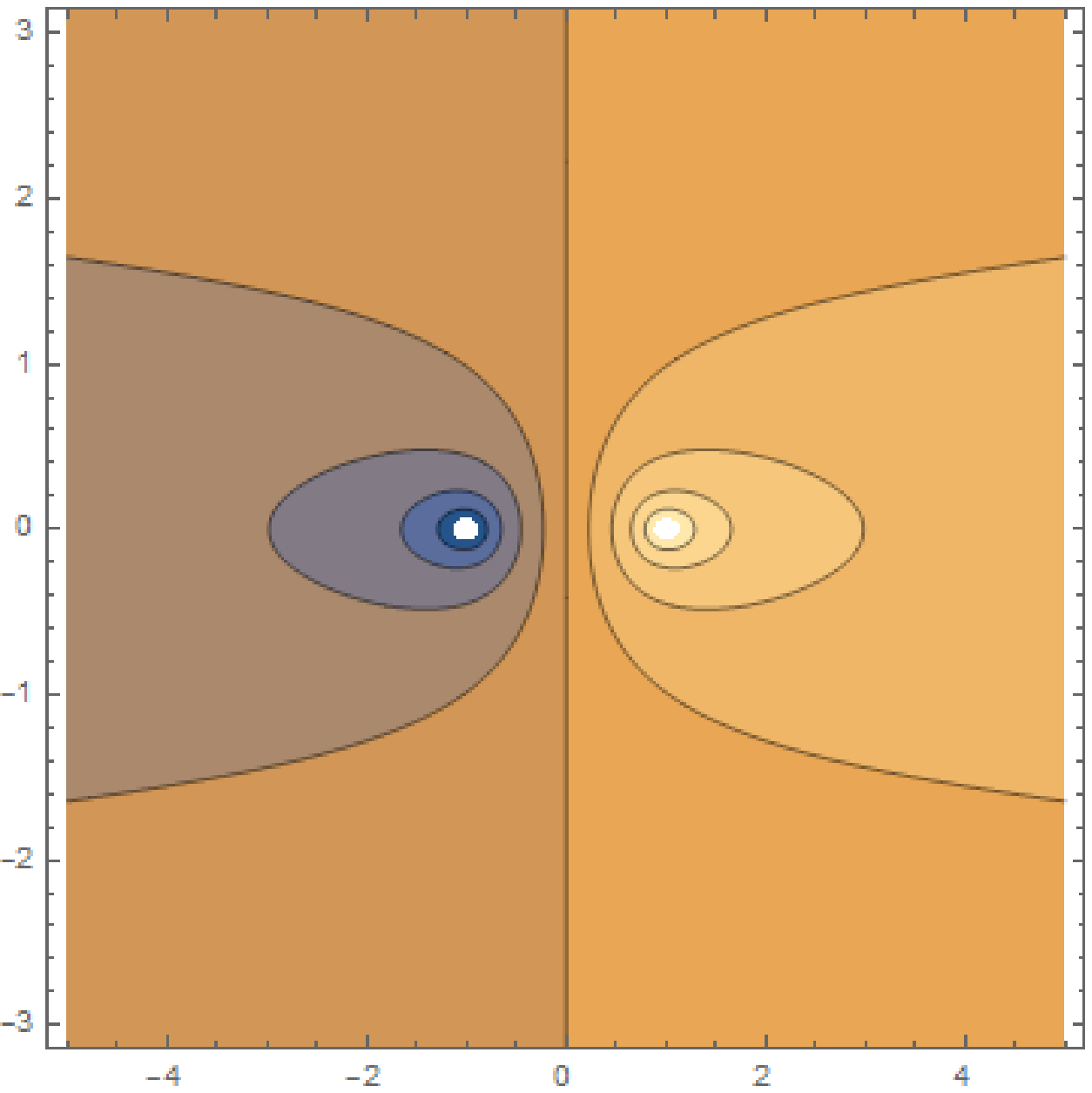}%
\\
Figure 13: $\ $Contour plot of $G_{o}$ for the squashed wormhole.
\end{center}}}

\end{center}

\end{document}